\documentstyle[12pt]{article}
\newcommand{\bq}{\begin{equation}}
\newcommand{\ee}{\end{equation}}
\newcommand{\fr}[2]{\frac{#1}{#2}}

\begin{document}
\pagestyle{plain}
\pagenumbering{arabic}

\begin{flushright}
{\large Preprint BUDKERINP 95-73}, September 1995\\
\end{flushright}

\begin{center}{\Large \bf Instanton--anti-instanton pair
induced contributions to $R_{e^+e^-\rightarrow hadrons}$ and
$R_{\tau \rightarrow hadrons}$}\\

\vspace{0.5cm}

{\bf S. V. Faleev,}\footnote{e-mail address:
S.V.Faleev@INP.NSK.SU}
{\bf  P. G. Silvestrov}\footnote{e-mail address:
P.G.Silvestrov@INP.NSK.SU}
 \\ Budker Institute of Nuclear
Physics, 630090 Novosibirsk, Russia

\vspace{0.5cm}

\end{center}
\begin{abstract}

The instanton--anti-instanton pair induced asymptotics of
perturbation theory expansion for the cross section of
electron--positron pair annihilation to hadrons and hadronic
width of $\tau$-lepton was found.  For $N_f = N_c$ the
nonperturbative instanton contribution is finite and may be
calculated without phenomenological input. The instanton induced
peturbative asymptotics was shown to be enhanced as $(n+10)!$
and in the intermediate region $n<15$ may exceed the renormalon
contribution.  Unfortunately, the analysis of  $\sim 1/n$
corrections shows that for $n \sim 10$ the obtained asymptotic
expressions are at best only the order of magnitude estimate.
The asymptotic series for $R_{e^+ e^- \rightarrow hadrons}$ ,
though obtained formally for $N_f =N_c$, is valid up to energies
$\sim 15$Gev. The instanton--anti-instanton pair nonperturbative
contribution to $R_{\tau \rightarrow hadrons}$ blows up. On the
one hand, this means that instantons could not be considered
{\it ab--initio} at such energies. On the other hand, this
result casts a strong doubt upon the possibility to determine
the $\alpha_s$ from the $\tau$--lepton width.

\end{abstract}


\section{Introduction. Instanton -- Renormalon.}\label{sec:0}

The semiclassical technique for  high order perturbation theory
estimates was suggested by L.N.Lipatov \cite{Lipatov} almost 20
years ago.  The main advantage of this approach is that there is
no need to deal with the Feynman graphs in order to find the
asymptotics. Only the contribution of one "the most important"
classical configuration in the functional space (instanton) is
considered, which can be associated with the sum of a huge
number of these graphs.

Shortly after the Lipatov's paper was published another
mechanism for the factorial growth of the series of perturbation
theory was found \cite{t'Hooft,ren}.  It turns  out that in the
theories with the running coupling constant a very simple chain
of graphs may dominate in the asymptotics of perturbation
theory. By analogy with the instanton this chain of graphs was
called renormalon.

In the present work we will calculate the instanton
induced contribution to the asymptotics of perturbation theory
series for the cross section of electron--positron  pair
annihilation and hadronic width of $\tau$--lepton.  As we shell
see, at $n<15$ the instanton contribution to the $n$-th order
may be quite competitive with the renormalon contribution.

During the last few years the issue of the asymptotic behaviour
of perturbation theory in QCD and QED (see e.g.
\cite{West}-\cite{Broa}) has attached the renewed interest.
However,  the  main  attention  was paid to the renormalon
asymptotics.

There are two types of renormalons -- ultraviolet and infrared
\cite{t'Hooft}. The usual form of the ultraviolet renormalon
contribution to, for example, $R_{e^+e^-\rightarrow hadrons}$
is:
\bq\label{uren}
R_{e^+e^- \rightarrow hadrons} = \sum R_n \left(
\fr{\alpha_s}{4\pi}\right)^n \,\,\, , \,\,\, R_n\sim \left(-b
\right)^n n!\ ,
\ee
where $b =\fr{11}{3} N_c -\fr{2}{3} N_f\approx 10$. In QCD the
series (\ref{uren}) is sign alternating and at least the Borel
sum of the series is well defined. The problems with summation
of ultraviolet renormalon appear in QED, where the coefficient
$b$ is negative.

Generally speaking, the infrared renormalon contribution to the
asymptotics depends on the process considered. For
$R_{e^+e^-\rightarrow hadrons}$ and $R_{\tau \rightarrow
hadrons}$ the contribution of the first infrared renormalon to
the asymptotics reads:
\bq\label{iren}
R_n \sim \left( \fr{b}{2} \right)^n
n! \,\,\, .
\ee

All problems associated with the asymptotic series become more
clear if the physical quantities are presented in the form of
Borel integral:
\bq\label{Bor}
R_{e^+e^-}= \int_0^{\infty} \exp \left( - \fr{4\pi}{\alpha}
t\right) F(t) dt \,\,\, .
\ee
Now the singularities of the function $F(t)$ on the complex
plane are the natural sources of asymptotic series.  The first
ultraviolet renormalon in QCD (\ref{uren}) corresponds to the
singularity at negative value $t = -1/b$. The first singularity
associated with infrared renormalon (\ref{iren}) is located
twice father from the origin, but at positive value $t = 2/b$.
Thus, the terms of the series (\ref{iren}) grow with $n$ more
slowly than those of the series (\ref{uren}). On the other hand,
for the infrared renormalon one has to choose the correct way
of the integration over singularity which lies on the positive
real axis. Different definitions of the integral (\ref{Bor})
will lead to the different power corrections $\delta R \sim
1/q^4$.

The issue of infrared renormalon is directly related with the well
known problem of Landau pole in theories with running coupling.
The singularity of Borel transform  $F(t)$ (\ref{Bor})
associated  with  infrared renormalon coincides with Landau pole
which  arises when the exchange of soft gluon in fermion loop is
taken into account (see the example of renormalon type graphs in
any of the papers \cite{t'Hooft}-\cite{Vainshtain}).  At this
point one may say that the main unsolved problems of QCD
(confinement) are also hidden in the infrared renormalon
\footnote{In fact, the situation with ultraviolet renormalon is
also not very clear.  Being written more accurately than
(\ref{uren}), the ultraviolet  renormalon contribution takes the
form:
\bq\label{renuv}
R_n=An^{\gamma}\left(-b\right)^n n!\ ,
\ee
where the constant $\gamma \sim 1$. It has been believed for a
long time that the main contribution to the ultraviolet
renormalon comes from the graphs with exchanging of one dressed
gluonic line as in the case of infrared renormalon. However, in
the recent work \cite{Vainshtain}  Vainshtain and Zakharov
showed, that the contribution of these graphs is small compared
to the graphs with two, three etc. dressed lines.  As a
result, only the constant $\gamma$ is known now and it is
absolutely unclear how to search for the overall factor $A$.}.

As we have already said, another approach for high order
perturbation theory estimates based on the consideration of the
specific ("classical") large fluctuations in the functional
space was suggested by Lipatov. The natural example of such
important fluctuations in QCD is the instanton--anti-instanton
pair. The generic form of instanton induced asymptotics appears
to be:
\bq\label{insas}
R_{nIA}\sim  (n+4N_c)!
\,\,\,
\ee
(here we do not show either overall numerical factor, which
also may be sufficiently large, or the factor $n^{\gamma}$ with
$\gamma \sim 1$). On the Borel plane the instanton -
anti-instanton pair corresponds to the pole at rather distant
point $t=1$, but residue in the pole turns out to be very large
$\sim (4\pi/\alpha_s)^{4N_c}$.

Thus, we can see that though the renormalons
(\ref{uren},\ref{iren}) do dominate at very large $n$ ($n>15$),
the instanton-induced contribution may dominate in the
intermediate asymptotics $n= 5\div 15$. If so, the pure
renormalon behavior (\ref{uren}) will never be observed in
directly calculated terms of perturbation theory due to a strong
competition with the instanton contribution.

It is clear, that the exactly known $3\div4$ terms of
perturbation theory series (for $\beta$--function,
$R_{e^+e^-\rightarrow hadrons}$ or $R_{\tau \rightarrow
hadrons}$) are much smaller than the estimate (\ref{insas}) and
the question at what number $n$ the perturbative series could
reach the full strength (\ref{insas}) is open now.  The only way
to answer this question is to consider the $\sim 1/n$
corrections to the leading asymptotics.  In this paper we have
found (and summed up) two subseries of corrections to $R_{nIA}$.
The first one behaves like $(N_c^2\ln{(n)}/n)^k$, the second -
like $(N_c^2\ln{(N_c)}/n)^k$.  The final expressions for
$R_{nIA}$ are given below (\ref{asymR}), (\ref{Rntau}). The
$\sim 1/n$ corrections tend to decrease the asymptotic
predictions and thus improve the agreement with the
"experiment". But important is not this agreement.
Unfortunately, the corrections look like $\sim N_c^2/n$ and,
thus, even for $n\sim 10$  our asymptotic expression can be used
at best as the estimate of the order of magnitude.

Up to now the instanton induced asymptotics of perturbation
theory for $R_{e^+e^- \rightarrow hadrons}$ was considered only
in the paper of I.I.Balitsky \cite{Balitsky}. Many of the
technical methods used by Balitsky were useful in our work, but
nevertheless, he could not find correctly either asymptotics of
perturbation theory, or nonperturbative contribution of the
instanton - anti-instanton pair in the most actual case
$N_f=N_c$.  The instanton induced contribution to the
perturbative asymptotics for the simple correlator of two
gluonic currents $j \sim \alpha_s [G^a_{\mu\nu}]^2$  has also
been considered by one of the authors of the present paper in
recent work \cite{Silvestrov}.

Similarly to the infrared renormalon (\ref{iren}) all terms of
the series (\ref{insas}) are of the same sign.  Nevertheless,
the problem of summing the series (\ref{insas}) seems not as
hopeless as that of the renormalon. Following G.'t Hooft
\cite{t'Hooft},  the author of \cite{Balitsky}  proposed to
rewrite the integral over the instanton--anti-instanton pair in
the Borel form by considering the action as a collective
variable. Within this approach the well--separated
instanton--anti-instanton pair is responsible for the singular
part of Borel function, while the ambiguous strongly interacting
instanton and anti-instanton contribute to its smooth part. On
the other hand, the best way to describe the smooth part of the
Borel transform is to calculate exactly the few first terms of
perturbative expansion. The accurate subtraction from the
singularity of dilute gas contribution in the toy model (double
well oscillator) allowed us to find the finite nonperturbative
instanton--anti-instanton contribution \cite{Faleev}.  In QCD at
$N_f=N_c$ the Borel integral diverges only logarithmically and
the total nonperturbative contribution from
instanton--anti-instanton pair may be found by cutting the
instanton size at $\rho \ll 1/\Lambda_{QCD}$.

In the next section we combine some useful results concerning
the instanton--anti\-ins\-tan\-ton pair and the behavior of
light fermions in instanton background. The correlation function
of two electro-magnetic currents is calculated in section {\bf
3}.  At $N_f=N_c$ we calculate the finite nonperturbative
instanton contribution as well as the instanton induced
asymptotics of the perturbation theory. The issue of $\sim 1/n$
corrections to the asymptotics is also considered.
Unfortunately, as we have already mentioned, the asymptotic
expression for the instanton contribution may be considered at
best only as the estimate of the order of magnitude up to  $n
\sim 10$.  On the other hand if $n$ is not very large the
instanton may make completely unvisible the renormalon
contribution. The possibility, discussed in section {\bf 4}, to
extrapolate our asymptotic formulas for $R_{e^+e^- \rightarrow
hadrons}$, found for $N_f = N_c$, up to energy $\sim 10-15 Gev$
may be considered as a poor consolation. At least at such
energies the high orders of the perturbation theory are not lost
at the background of standard power corrections. More important
is the issue of instanton contribution to $R_{\tau \rightarrow
hadrons}$ (section {\bf 5}) due to numerous attempts to extract
from this quantity the value of $\alpha_s(m_{\tau})$.
Unfortunately, here our result is negative too.  Even if one
does not take into account the asymptotics of the perturbation
theory (which also has to be reached), the pure nonperturbative
contribution to $R_{\tau \rightarrow hadrons}$ turns out to be
one or two orders of magnitude larger than one can admit. In
spite of the fact, that at energy $m_{\tau}c^2 \sim 2Gev$ our
result is again only the estimate of the order of magnitude, we
have a strong doubt about the possibility to find the accurate
value of $\alpha_s$ from the $\tau$--lepton width.

\section{The Instanton--anti-instanton pair}\label{sec:0.5}

The interesting physical quantities, like, e.g.,
$R_{e^+e^-\rightarrow hadrons}$, may be naturally found via the
analytical continuation of the corresponding Euclidean
correlators:
\begin{eqnarray}\label{Pi}
R_{e^+e^-\rightarrow hadrons}(q^2)&=&12\pi Im\ \Pi
(-q^2-i\varepsilon )\
\ ,
\\
\Pi_{\mu\nu}(q^2)=\Pi(q^2)(q_{\mu}q_{\nu}-q^2\delta_{\mu\nu})&=&
\int d^4 x e^{iqx}\langle j_{\mu}(x)j_{\nu}(0)\rangle \ \ ,
\nonumber
\end{eqnarray}
where $j_{\mu}=\sum_{flavors}e_f\Psi^+_f\gamma_{\mu}\Psi_f$.
Therefore, we have to calculate the contribution of the
instanton--anti-instanton pair to the correlator.  As we have
said above, the strongly interacting instanton and
anti-instanton contribute only to the smooth part of the Borel
function  $F$ (\ref{Bor}) and do not effect the asymptotics
of perturbation theory. The singular part of the Borel function
is saturated by the almost non-interacting pseudoparticles.  That
is why, one can obtain the reliable prediction for the
asymptotics starting from such ill defined object as the
instanton--anti-instanton pair. More concretely, the field
configuration relevant for the large orders of perturbation
theory is a small instanton inside of a very large
anti-instanton (or vice versa). The size of small instanton is
regulated by the internal momentum in correlator (\ref{Pi}) $q
\rho_I \sim 1$.  The size of anti-instanton (as well as the
distance between the centers of pseudoparticles $R \sim \rho_A
\gg \rho_I$) determines, how close we are to the singularity on
the Borel plane.

We are interested in the instanton--anti-instanton interaction
in the leading approximation. Therefore, the simple
sum of instanton and anti-instanton may be used:
\bq\label{ansatz}
A_{\mu} = U_A A_{\mu}^A U_A^+ + U_I A_{\mu}^I U_I^+ \,\,\, ,
\ee
where $U_A, U_I$ are the constant $SU(N_c)$ matrices in the
color space.  By trivial gauge rotation one can make $U_I=1$.
For "small" instanton the singular gauge seems to be preferable:
\bq\label{Ains}
A_{\mu}^I = \fr{\overline{\eta}_{\mu\nu}(x-x_I)_{\nu} \rho_I^2}
{(x-x_I)^2((x-x_I)^2+\rho_I^2)} \,\,\, ,
\ee
where $\overline{\eta}_{\mu\nu}\equiv \tau^a
\overline{\eta}^a_{\mu\nu}$, and $\tau^a$ are the usual Pauly
matrices located in the upper left  $2 \times 2$  corner
(another elements vanish) of $N \times N$ matrix describing the
gluonic fields.

Before we add the anti-instanton to (\ref{Ains}), the singularity
at $x=x_I$ is pure gauge singularity.  In order to suppress the
unphysical singularities in the sum (\ref{ansatz}), one may
choose $A_{\mu}^A$ in any regular gauge which leads to
$A_{\mu}^A(x=x_I)=0$. For example, one may slightly rotate the
BPST anti-instanton:
\bq\label{Aans}
A_{\mu}^A = S\left[ \fr{\overline{\eta}_{\mu\nu}(x-x_A)_{\nu}}
{(x-x_A)^2+\rho_A^2}
\right] S^+ + i S\partial_{\mu} S^+\,\,\, .
\ee
Near the center of instanton the matrix $S(x)$ satisfying the
above condition has the form:
\bq\label{SSS}
S= e^{i\Theta} \,\,\,\, , \,\,\,\, \Theta = B_{\mu} (x-x_I)_{\mu}
+C_{\mu\nu}(x-x_I)_{\mu}(x-x_I)_{\nu} \,\,\,\, ,
\ee
where $B_{\mu} = - \fr{\overline{\eta}_{\mu\nu}
R_{\nu}}{R^2+\rho_A^2} = -
A_{\mu}^A(x=x_I)$ , $R_{\mu}= (x_A-x_I)_{\mu}$ and $C_{\mu\nu} =
C_{\nu\mu}$ is an arbitrary symmetric tensor.

After direct calculation the classical action of the
instanton--anti-instanton configuration may be found with the
usual dipole--dipole interaction of pseudoparticles:
\bq\label{action}
S_{IA} = \fr{4\pi}{\alpha_s}\left\{
1 - \xi h\right\} \,\,\,\, , \,\,\,
\xi = \fr{\rho_I^2 \rho_A^2}{(R^2 + \rho_A^2)^2} \,\,\, ,
\,\,\,
h=2|TrO|^2 -TrOO^+ \,\,\, ,
\ee
and $O$ is the upper left $2 \times 2$ corner of the matrix
$U=U_A^+ U_I$ (\ref{ansatz}).

The features of light fermions are mostly sensitive to the
presence of instantons. The Dirac operator $\hat{D}$ for each
flavor of massless fermions has two eigenfunctions $\Psi_{\pm}$
with anomalously small eigenvalues $\lambda_{\pm}$. As we
consider the case of almost noninteracting pseudoparticles it is
natural to search for these eigenfunction in the form of linear
combination of nonperturbed zero modes of separate
pseudoparticles. Explicit expressions for zero modes in the
background of singular instanton and regular anti-instanton are:
\begin{eqnarray}\label{Psi0}
\Psi_I=\fr{1}{\pi}\fr{\rho_I}{[x^2+\rho_I^2]^{3/2}}
\fr{x_\mu\tau^-_\mu}
{|x|} \left( \begin{array}{c} \phi \\ \phi \end{array}\right)
 \,\,\, , \,\,\,
\Psi_A=\fr{1}{\pi}
\fr{\rho_A}{[x^2+\rho_A^2]^{3/2}} U \left( \begin{array}{c}
\phi
\\ -\phi
\end{array}\right)
 \,\,\, ,
\end{eqnarray}
where $\phi^{\alpha m}=\varepsilon_{\alpha m}/\sqrt{2}$ for
$\alpha=1,2$ and $\phi^{\alpha m}=0$ for $\alpha > 2$, $\alpha$
is color index, $m=1,2$ is spinor index, $\varepsilon_{\alpha
m}$ is an antisymmetric tensor, and $\tau^{\pm}=(\mp
i,\stackrel{\rightarrow}{\tau})$.  All correlation functions
which are of interest for us are saturated by the region
$|x-x_I| \sim \rho_I$. Within this region the anti-instanton
zero mode should be modified.  It is easy to verify that the
spinor function
\begin{eqnarray}\label{PsiA}
\Psi_A=\fr{1}{\pi} \sqrt{ \fr{(x-x_I)^2}{(x-x_I)^2+\rho_I^2}
} \,
\fr{\rho_A}{[R^2+\rho_A^2]^{3/2}} U \left( \begin{array}{c} \phi
\\ -\phi
\end{array}\right)
 \,\,\, ,
\end{eqnarray}
is the solution of the Dirac equation  $\hat{D}_I
\Psi_A =0$ at $x-x_I \sim \rho_I$ and approaches  (\ref{Psi0}) at
$|x-x_I| \gg \rho_I$.

After diagonalization of the Dirac operator $\hat{D}_{I+A}$
within the subspace of zero modes (also the identity $\lambda_+
\equiv - \lambda_-$ may be useful) one gets:
\bq\label{lambda}
\lambda_{\pm}= \pm \fr{2\rho_I \rho_A}{(\rho_A^2 +
R^2)^{3/2}}|TrO| \,\,\, \ \ ,
\Psi_{\pm}=\fr{1}{\sqrt{2}}\bigg(\Psi_I \pm
\fr{TrO^+}{|TrO|} \Psi_A\bigg)
\,\,\, .
\ee

Because the instantons interact very slightly the nonzero modes
contribution to the fermionic determinant factorizes. The Green
function in this case also has rather simple form:
\bq\label{SSum}
S(x,y)=S_{\lambda}+G_I+G_A-G_0+O(\xi)
\,\,\, .
\ee
Here, $G_0(x-y)$ is the bare Green function, $G_I,G_A$ are Green
functions in the background of separate instanton and
anti-instanton and $S_{\lambda}$ is the zero mode contribution.
{}From (\ref{Psi0}-\ref{lambda}) one gets:
\bq\label{GF}
S_{\lambda}(x,y)=\fr{\Psi_+(x)\Psi_+^+(y)}{\lambda_+}
+\fr{\Psi_-(x)\Psi_-^+(y)}{\lambda_-}
= \fr{1}{|\lambda|}\left\{ \fr{TrO^+}{|TrO|} \Psi_A \Psi_I^+
+\Psi_I \Psi_A^+  \fr{TrO}{|TrO|} \right\} \,\,\, .
\ee
The Green function $S(x,y)$ may contain only the terms which
convert the right fermions to the left and vice versa. That is
why the largest, proportional to $\Psi_I \Psi_I^+$, term in
$S_{\lambda}$ vanishes. As a result, the contributions of the
zero modes and of "quantum" Green function $G_I$ to the
correlation functions are of the same order of magnitude.  The
last two terms in (\ref{SSum}) almost cancel $G_A - G_0
\sim \xi$ in the region of interest $|x-x_I|\sim \rho_I \ll
\rho_A$.  The Green function  in the instanton field (for
simplicity, we put  $x_I=0$) has been found in \cite{Brown}:
\begin{eqnarray}\label{GI}
2\pi^2 G_I(x,y)\! &=& \! \fr{i\gamma_{\mu}}{\sqrt{T_x T_y}}
\fr{(x\tau^-)}{|x|}\bigg[  z_{\mu} \fr{\rho^2+(\tau^+ x)(\tau^- y)}
{z^4} + \tau^+_{\mu}\fr{(z\tau^-)\rho^2}
{2z^2 T_x} \bigg] \fr{(\tau^+ y)}{|y|} \bigg(\fr{1+\gamma_5}{2}\bigg)
\nonumber \\
 &\,&
 +(\,c.c.,x\leftrightarrow y)  \ \, ,
\end{eqnarray}
where $T_x=\rho^2+x^2$ and $z=x-y$.
We use the Hermitean euclidean matrices $\gamma_{\mu}$ which
satisfy $\{ \gamma_{\mu}, \gamma_{\nu}\} = 2\delta_{\mu\nu}$
and, for example, the bare Green function satisfies the equation
$-i\gamma_{\mu}\partial_{\mu} G_0 = \delta (x-y)$.

\section{The calculation of the correlation function.}\label{sec:1}

Now, at last, we can write down the expression for the
correlation function (\ref{Pi}):
\begin{eqnarray}\label{pol.op}
\Pi_{\mu\nu}=\!&2&\!\!\! \int e^{iqx}
\exp{\left\{ \fr{4\pi}{\alpha_s}\xi h\right\}}
\bigg\{-\sum\limits_f e_f^2 \, T_{\mu \nu}(x,0)+
\bigl( \sum\limits_{f} e_f  \bigr)^2
B_{\mu}(x) B_{\nu}(0)\bigg\}\times
\nonumber \\ &\,&[4 \xi^{3/2}
|TrO|^2]^{N_f} \fr{d(\rho_I)}{\rho_I^5}
\fr{d(\rho_A)}{\rho_A^5} dx dx_I dx_A d\rho_I d\rho_A dU \ \, ,
\end{eqnarray}
where
\begin{eqnarray}\label{Tmunu}
T_{\mu\nu}(x,y) &=& Tr\bigl\{ \gamma_{\mu} S(x,y) \gamma_{\nu}
S(y,x) \bigr\} - Tr\bigl\{ \gamma_{\mu} G_0(x,y) \gamma_{\nu}
G_0(y,x) \bigr\} \,\,\, , \\ &\,&
B_{\mu}(x) = Tr\bigl\{ \gamma_{\mu} S(x,x) \bigr\} \,\,\, .
\nonumber
\end{eqnarray}
The factor $2$ in front of the integral in (\ref{pol.op})
accounts for the equal contribution from small anti-instanton
and large instanton.  The factor  $\sim \xi^{3N_f/2}
|TrO|^{2N_f}$ in the square brackets accounts for the
contribution of almost zero modes  (\ref{lambda}) to the fermion
determinant.  The instanton density reads
\cite{t'Hooft76,Bernard,Hasenfratz}:
\begin{eqnarray}\label{eq:dins}
d(\rho)= \fr{c_1e^{-N_cc_2+N_fc_3}}{(N_c-1)!(N_c-2)!}
\left( \fr{2\pi}{\alpha_s(\rho)}\right)^{2N_c}
\exp\left(-\fr{2\pi}{\alpha_s(\rho)} \right)  \,\,\, .
\end{eqnarray}
for $\overline{MS}$ scheme $c_1=2e^{5/6}/\pi^2$, $c_2=1.511$,
$c_3=0.292$, $c_2-c_3=2\ln{2}-1/6$.  Up to now, in many papers
the wrong values of $c_2$ and $c_3$ have been used (for which,
in particular, $c_2-c_3=2\ln{2}$), though the error in
\cite{t'Hooft76} which had been done while passing from
Pauli--Villars to dimensional regularization was corrected in
\cite{Hasenfratz} and t'Hooft in later papers used the correct
expression for $d(\rho)$ .

We will also use the well known two--loop formula:
\bq\label{eq:alpha}
\fr{4\pi}{\alpha_s(p)} = b\ln
\left(\fr{p^2}{\Lambda^2}\right)+
\fr{b'}{b}\ln\left(
\ln\left(\fr{p^2}{\Lambda^2}\right)\right) + ...
\,\,\, ,
\ee
where $b=\fr{11}{3} N_c -\fr{2}{3} N_f$ and $b'=\fr{34}{3} N_c^2
- \fr{13}{3}N_f N_c + N_f/N_c $. If $N_f=N_c=3 $ one has $b=9$
and $b'=64$.

In order to reduce the integral (\ref{pol.op}) to the Borel one,
we have to integrate over all collective variables $x_I, x_A,
\rho_I, \rho_A, U$, and, also, over  $x$ at fixed action
(\ref{action}), that is to say, at fixed value  of the
combination $\xi h$. This problem may be divided into two parts.
The integration over $x$ and $x_I$ is rather tedious algebraical
problem due to the complicated form of the correlation function
of quark currents in the instanton field
(\ref{pol.op}),(\ref{Tmunu}),(\ref{GI}). But from physical point
of view, the main problem is the integration over size $\rho_A$
and position $R=x_A-x_I$ of the large anti-instanton.  There are
two competing effects here. First, the factor $d(\rho_A) \sim
\rho_A^{b}$ tends to make the integral over $\rho_A$ divergent.
Second, the almost zero fermion modes $\lambda^{2N_f} \sim
\rho_I^{2N_f}/\rho_A^{4N_f}$ (\ref{lambda}) tend to suppress the
contribution of large anti-instantons.  As a result, the value
of the integral (\ref{pol.op}) (as well as the validity of our
method) depends strongly on the number of light quarks $N_f$.
The simple dimensional analysis (let us note, that the
integrations over $d\rho_A$ and $d^4x_A$ are not independent and
$\rho_A \sim x_A-x_I$ owing to the constraint $\xi h = const$
(\ref{action})) shows that the critical value is $N_f = N_c$.
If $N_f< N_c$ the first effect dominates and the integral
(\ref{pol.op}) diverges at large $\rho_A$. Nevertheless, just in
this case the well defined instanton induced asymptotics of
perturbation theory may be extracted from (\ref{pol.op}). For
calculation of the integral (\ref{pol.op}) beyond the
perturbation theory at $N_f< N_c$ the new physical income is
necessary (for example, one may consider the instanton liquid).
The most favorable case is $N_f = N_c$. In this case, the
integral over $\rho_A$ in $(\ref{pol.op})$ diverges only
logarithmically. As a result, we are able not only to obtain the
asymptotics of perturbation theory, but also to calculate (at
least with the logarithmic accuracy) the finite nonperturbative
instanton--anti-instanton pair contribution to
$R_{e^+e^-\rightarrow hadrons}$ and $R_{\tau \rightarrow
hadrons}$.

If $N_f > N_c$, the attraction of pseudoparticles which appears
owing to fermionic zero modes prevails. As a result, the
integral is saturated by $\rho_A \sim R \sim \rho_I$ and the
approximation of almost noninteracting pseudoparticles does not
work.  The problem of instanton induced asymptotics might also
have the solution in this case, but, at least, this solution
requires the considerable modification of the method used in the
present paper.

With the use of (\ref{eq:dins}), (\ref{eq:alpha}) let us extract
the $\rho_A$ dependent part from (\ref{pol.op}):
\bq\label{eq:phi}
d(\rho_A) = \phi (\rho_I^2 / \rho_A^2 )
 \left( \fr{\rho_I}{\rho_A}
\right)^b d(\rho) \,\,\,\, , \,\,\,\,
\phi(x) = \left[ 1 +
b\fr{\alpha_s}{4\pi}\ln (x)\right]^{2N_c-\fr{b'}{2b}}
\ee
Here and below $\alpha_s \equiv \alpha_s(q) \simeq \alpha_s
(\rho_I)$. In order to find the leading perturbative
asymptotics, one may assume $\phi(x)\equiv 1$ (as it was done in
\cite{Balitsky}), but the calculation of the correlator
(\ref{pol.op}) beyond the perturbation theory requires the use
of the function $\phi(x)$ in the form (\ref{eq:phi}).  Moreover,
the expansion of $\phi(x)$ in a series in $(\alpha_s
\log(x))^k$ generates an important set of preasymptotic
corrections $\sim (\log(n)/n)^k$ to the leading asymptotics.

The integral over $\rho_A$ and $R=x_A-x_I$ for a fixed value of
$\xi$ gives:
\bq\label{eq:rhoA}
\int \phi (\rho^2 / \rho_A^2 ) \rho_A^{b-5} \delta \left(
\fr{\rho^2\rho_A^2}{(R^2+\rho_A^2)^2}-\xi\right) d\rho_A
d^4R =
\fr{\pi^2}{2(b-2)(b-1)} \fr{\rho^b}{\xi^{b/2+1}}
\phi(\xi) \,\,\, .
\ee

Now we would like to present in more compact form the
expressions for $T_{\mu\nu}$ and $B_{\mu}$ (\ref{pol.op}).
Actually, the traces $T_{\mu\mu}$ and $B_{\mu}(x)B_{\mu}(y)$ are
of interest because the polarization operator $\Pi_{\mu\nu}$ is
transverse. As for $B_{\mu}$, the situation is rather simple.
Formally, the Green function at coincident points involved to
$B_{\mu}$ goes to infinity.  Nevertheless, the contribution to
$B_{\mu}$ from the one-instanton Green function (\ref{GI}) has
to be of the form:
\bq\label{Bf}
Tr \{ \gamma_{\mu} G_I(x,x) \} = (x-x_I)_{\mu} f(x-x_I)
\ee
because $(x-x_I)_{\mu}$ is the only possible vector. As a
consequence of transversness of $B_{\mu}$: $\partial_{\mu}
B_{\mu}=0$, one immediately gets $f(x)\equiv 0$. Thus, the only
nontrivial contribution to $B_{\mu}$ originates from zero modes:
\begin{eqnarray}\label{Bmumu}
B_{\mu}(x)B_{\mu}(y) &=& Tr\{ \gamma_{\mu} S_{\lambda}(x,x)\}
Tr\{ \gamma_{\mu} S_{\lambda}(y,y)\} =
\fr{(x,y)}{\pi^4 T_x^2 T_y^2} (H-2D) \,\,\,\, , \\
H\!&=&\! \fr{TrO^+O}{|TrO|^2} \,\,\,\, , \,\,\,\,
D= Re \left( \fr{DetO}{[TrO]^2} \right)
 \,\,\,\,\, . \nonumber
\end{eqnarray}
Here $x_I =0$ .

After some tedious algebraic manipulations one gets:
\begin{eqnarray}\label{Tmumu}
T_{\mu\mu}(x,y) \!&=&\! \fr{1}{\pi^4 T_x^2 T_y^2} \left\{
\fr{2\rho^4}{z^2}
- \fr{4\rho^2 (r,z)^2}{ z^4} +
\rho^2 - 2(x,y) D
\right\} \\
&\,& - \fr{1}{\pi^4 z^4}
\left\{
\fr{(y,z)}{T_y^2} - \fr{(x,z)}{T_x^2} \right\}
 +\left( \begin{array}{c} \mbox{odd over}\,\,\, r_{\mu}\\
\mbox{terms} \end{array} \right) \,\,\,\,\, ,
\nonumber
\end{eqnarray}
where $z_{\mu}=(x-y)_{\mu}$, $r_{\mu} = \fr{1}{2} (x+y)_{\mu}$ .
Here the coordinates  $x$ and $y$ are measured from the
instanton center and hence, like it was done in the papers
\cite{Gross,JEllis}, the integral over $d^4x_I$ may be replaced by the
integral over $d^4r$
\begin{eqnarray}\label{TBR}
\!&\,& \int T_{\mu\mu}(x,y) d^4r = \fr{1}{2\pi^2 z^2}+
\fr{2}{\pi^2} \int_0^1 du \fr{u\overline{u}}{T^2} \left\{
u\overline{u}\rho^2 - ( T +\rho^2)D
\right\} \, , \\
\!&\,& \int B_{\mu}(x)B_{\mu}(y) d^4r =
\fr{1}{\pi^2} \int_0^1 du
\, u\overline{u}  \fr{T+\rho^2}{T^2}
(H-2D)
\,\,\,\,\, , \nonumber
\end{eqnarray}
where $T=\rho^2 + u\overline{u} \, z^2$ and $\overline{u} = 1-u$ .
Now it is easy to find the Fourier transforms:
\begin{eqnarray}\label{TBq}
\!&\,& \int e^{iqx}T_{\mu\mu}(x,0) d^4x d^4x_I = \fr{2}{q^2}+
4\rho^2 \int_0^1 du  \left\{
K_0\left(\fr{q\rho}{\sqrt{u\overline{u}}}\right) -
\fr{1}{u\overline{u}}
K_2\left(\fr{q\rho}{\sqrt{u\overline{u}}}\right) D
 \right\} \, , \\
\!&\,& \int e^{iqx} B_{\mu}(x)B_{\mu}(0) d^4x d^4x_I =
2\rho^2 \int_0^1 \fr{du}{u\overline{u}}
K_2\left(\fr{q\rho}{\sqrt{u\overline{u}}}\right)
(H-2D)
\,\,\,\,\, , \nonumber
\end{eqnarray}
where $K_0$ and $K_2$ are the McDonald functions.

The last integral we are to calculate is the integral over the
size of the "small" instanton -- $\rho$. As we have seen above
(\ref{Pi},\ref{eq:rhoA}), the $\rho_I$-dependence of the
integrand (except for that contained in (\ref{TBq})) in the
large logarithms approximation arises as the multiplier
$d^2(\rho)/\rho^5 \sim \rho^{2b-5}$ . From the point of view of
the Operator Product Expansion the large instanton and
anti-instanton lead to the power correction in $\Pi_{\mu\nu}$
\bq\label{q2}
\fr{(q_{\mu}q_{\nu} - q^2\delta_{\mu\nu} )}{q^4}
\fr{\alpha_s}
{6\pi}\sum\limits _f
e_f^2 \int d^4x
Tr\bigg(G_{\mu\nu}^{I+A}(x)G_{\mu\nu}^{I+A}(x)\bigg)\ .
\ee
In our formulas this power correction corresponds to the term
$2/q^2$ in the first integral (\ref{TBq}). The integral over the
instanton size for such correction to  $R_{e^+e^- \rightarrow
hadrons}$ diverges like $\sim1/q^4\int  d\rho_I \rho_I^{2b-5}$.
However, the  $\sim 1/q^4$ corrections, which account for the
long-wave vacuum fluctuations, are natural to be considered as
the part of the infrared renormalon.  Thus, we shall omit this
term.  All other terms in (\ref{TBq}) contain the exponentially
decreasing  McDonald functions and, hence, the $\rho$-integral
converges:
\begin{eqnarray}\label{Kint}
\int K_0 \rho^{2b-3} d\rho du = \fr{2^{2b-4}}{q^{2b-2}}
\fr{[(b-1)!(b-2)!]^2}{(2b-1)!} \,\,\,\, , \,\,\,\, \\
\int K_2 \rho^{2b-3} \fr{d\rho du}{u\overline{u}} =
\fr{2^{2b-4}}{q^{2b-2}}
\fr{(b-1)![(b-2)!]^2(b-3)!}{(2b-3)!} \,\,\, . \nonumber
\end{eqnarray}
The coefficient  $b \sim 10$ may be considered as the large
parameter. Therefore, the both integrals (\ref{Kint}) have the
form  $\int e^{-2q\rho} \rho^{2b} d\rho$ and are saturated by
$\rho\approx b/q$~. Thus, the size of the "small" instanton
turns out to be $b$ times larger than the typical wave length.
In essence, both right hand sides of (\ref{Kint}) are $\sim
(b/q)^{2b} e^{-2b}$ .

At last, collecting together (\ref{Pi}), (\ref{pol.op}),
(\ref{TBq}) and (\ref{Kint}), we get:
\begin{eqnarray}\label{pol.op2}
\Pi(q^2)&=& -\fr{\pi^2}{3} 4^{b+N_f-2}\fr{[(b-2)!
(b-3)!]^2}{(2b-3)!}d^2(q)  \nonumber \\  &\times&
\int dU d\xi |TrO|^{2N_f} \xi^{\fr{11}{6}(N_f-N_c)-1}\phi(\xi)
 \exp{\left\{ \fr{4\pi}{\alpha_s}\xi h\right\}}
A \,\,\,\,\, ,
\end{eqnarray}
where (see (\ref{Bmumu}))
\bq\label{A&B}
A=\left( 4D - \fr{2b-4}{2b-1}\right) \sum e_f^2 +
(2H-4D) (\sum e_f)^2 \,\,\,\,\,.
\ee
This result coincides, up to misprints, with that obtained in
\cite{Balitsky} (of course for $\phi(\xi)=1$).

The formula for $\Pi(q^2)$ may be rewritten in the form of Borel
integral if one introduces the new variable $t=1-h\xi$:
\begin{eqnarray}\label{pol.op3}
\Pi(q^2)&=&Const \,
\left( \fr{4\pi}{\alpha_s(q)}\right)^{4N_c} \int dU
 |TrO|^{2N_f} A |h|^{\fr{11}{6}(N_c-N_f)} \times \nonumber \\
&\,&
 \left[ \theta(h) \int_0^1
(1-t)^{\fr{11}{6}(N_f-N_c)-1}
\exp{\left\{-\fr{4\pi}{\alpha_s}t\right\}}\phi \, dt+
\right. \nonumber \\
&\,& \,\, \left. \theta (-h) \int_1^{\infty}
  (t-1)^{\fr{11}{6}(N_f-N_c)-1}
 \exp{\left\{-\fr{4\pi}{\alpha_s}t\right\}}\phi \,
 dt\right] \,\,\, .
\end{eqnarray}
Here $\phi=\phi(|1-t|)$ (\ref{eq:phi}). It will be recalled that
formally $\phi = 1 + O(\alpha)$ , but close to the singularity
$t=1$ taking into account the function $\phi$ changes completely
the value of the integrand.  $\theta$-function in
(\ref{pol.op3}) takes the values $1$ or $0$ in accordance with
the sign of its argument.

As we have already said, our method does not work at $N_f >
N_c$~. In fact, if $N_f > N_c$ the integral (\ref{pol.op3}) over
the orientations of the matrix $U$ diverges at $h=0$. The
physical nature of the divergency is clear.  We have used the
Faddeev-Popov unity in order to fix the value of the
instanton--anti-instanton interaction   $S_{int}
=\fr{4\pi}{\alpha_s} (t-1)$ (recall that our formulas are valid
only for  $|t-1|\ll 1$).  However, if $N_f > N_c$ the main
contribution to the integral comes from the configurations for
which although the dipole-dipole interaction is formally small,
the pseudoparticles themselves have the comparable sizes and
strongly overlap. Of course, one can hardly find reliable result
in this situation. In particular, our method does not work for
such an attractive problem as the calculation of the
perturbation theory asymptotics for $\Gamma_{Z_0 \rightarrow
hadrons}$ ($N_c=3$, $N_f=5$).

Of course, our formulas are valid if $N_f < N_c$. However, there
are just $3$ light quarks in nature and the case $N_f < N_c$
seems to be mostly of academic interest. Furthermore, for $N_f <
N_c$ the correct result for $R_{e^+e^- \rightarrow hadrons}$ (up
to trivial modification of single instanton density, see
discussion after eq. (\ref{eq:dins})) has been obtained by
Balitsky  \cite{Balitsky}.

Therefore, we would like to consider in detail only the most
interesting case $N_f=N_c$. If, in addition, $N_c =3$ the
formula (\ref{pol.op3}) for  $\Pi(q^2)$ takes the form:
\begin{eqnarray}\label{Pi1}
\Pi(q^2)=-\fr{e^{8/3}[7!6!]^2}{3\pi^2 16!}
\left( \fr{4\pi}{\alpha_s}\right)^{12}
\left( 0.510
\int_0^1 dt\fr{\exp{(-\fr{4\pi}{\alpha_s}t)}}{1-t}\phi
+0.054 \int_1^{\infty} dt
\fr{\exp{(-\fr{4\pi}{\alpha_s}t)}}{t-1}\phi
 \right) \, ,
\end{eqnarray}
Here the averaging over $SU(3)$ group was performed numerically
(see Appendix).

The expression (\ref{Pi1}) is enough to find the leading asymptotics
of perturbative expansion for $\Pi (q)$
(here we put $\phi =1$):
\begin{eqnarray}\label{asymP}
\Pi(q^2)=\sum \Pi_{n} \left(\fr{\alpha_s}{4\pi}\right)^n
\,\,\, &,& \,\,\, \\
\Pi_{n} =- \fr{e^{8/3}}{3\pi^2} \fr{[7!6!]^2}{16!} 0.510
(n+11)!&=& - 0.156 (n+11)! \,\,\, . \nonumber
\end{eqnarray}
We see that instanton--anti-instanton induced contribution to
the  perturbation theory  series do has a huge enhancement
$(n+11)!$.

Both integrals in (\ref{Pi1}) diverge at $t=1$ (although the
main physical problem is to interpret the contribution of small
$t$ to $\Pi(q^2)$ (\ref{Pi1})).  In the configuration space
these divergences are related to the integration over almost
noninteracting instanton and anti-instanton. Since the
divergence is only logarithmic one can try out the physical
intuition in order to restrict the range of integration in
(\ref{Pi1}). The natural cut-off for $\rho_A$ is $\rho_A \ll
1/\Lambda_{QCD}$, or, in terms of $t$
\bq\label{eq:cut}
|t-1|_{min} \sim \left(\fr{\rho_I}{\rho_{Amax}}\right)^2
< \fr{\Lambda^2}{q^2} \,\,\, .
\ee
If so, the nonperturbative part of (\ref{Pi1}) may be found
explicitly.  To this end let us supplement the first term in
(\ref{Pi1}) up to the principal value integral:
\bq\label{Pint}
\Pi(q^2) \sim 0.510 \, P \!\! \int_0^{\infty}
dt\fr{\exp{(-\fr{4\pi}{\alpha_s}t)}}{1-t} \phi(|1-t|)+
\fr{48}{85} e^{-\fr{4\pi}{\alpha_s}}
\int_{(\Lambda/q)^2}^{\infty} \fr{dx}{x} \phi(x)
\exp{(-\fr{4\pi}{\alpha_s}x)}
\,\,\, .
\ee
By the use of the logarithmic variable $ y=\fr{b}{4\pi}
\ln(1/x)$ in the last integral one gets  (up to correction
$\sim\alpha_s$)
\bq\label{Inep}
\int_{(\Lambda/q)^2}^{\infty}\fr{dx}{x} \phi(x)
\exp{(-\fr{4\pi}{\alpha_s}x)} \approx \fr{4\pi}{b}
 \int_0^{1/\alpha_s}
(1- \alpha_s y)^{22/9} dy = \fr{4\pi}{31 \alpha_s}
\,\,\, .
\ee
Note, that if one replaces $\phi(x)$ by $1$, the result for the
nonperturbative correction will be 31/9 times larger.  Finally,
collecting together (\ref{Pint}) and (\ref{Inep}) one finds:
\bq\label{Pi2}
\Pi(q^2)=- \fr{e^{8/3}[7!6!]^2}{3\pi^2 16!}
\left( \fr{4\pi}{\alpha_s}\right)^{12}
\left( 0.510 \,
P\!\! \int_0^{\infty}
dt\fr{\exp{(-\fr{4\pi}{\alpha_s}t)}}{1-t} \phi(|1\!-t|)+
\fr{48}{2635} \fr{4\pi}{\alpha_s} e^{-\fr{4\pi}{\alpha_s}}
\right) \, .
\ee
Let us say a few words about the obtained nonperturbative
correction $\sim e^{-\fr{4\pi}{\alpha_s}}$.  Effectively, the
integration over $t$ in (\ref{pol.op3}), (\ref{Pi2}) may be
thought as the integration over the size of large
anti-instanton. In the logarithmic scale $\ln (\rho_A \Lambda)
\sim \alpha_s(\rho_A)^{-1} $ and one may say that the
integration is performed over the (inverted) running coupling
constant. The remarkable feature of our result  is that all the
values of $\alpha_s(\rho_A)$ in the whole range $\alpha_s
(\rho_I) < \alpha_s (\rho_A) \ll 1$ make the comparable
contribution to the nonperturbative part.

The formula (\ref{Pi2}) correctly accounts for the high orders
of the perturbation theory and the nonperturbative corrections.
However, the first terms of the perturbative expansion are
wrong in (\ref{Pi2}).  The only way to eliminate this
defect is to calculate explicitly the first $3 - 5 - 10$ terms
of the perturbation theory.  After such modification the
expression (\ref{Pi2}) takes the form:
\begin{eqnarray}\label{Pi3}
\Pi(q^2)&=& \sum_0^N \Pi_{n\, exact} \left( \fr{\alpha_s}{4\pi}
\right)^N - \\
&\,& \fr{e^{8/3}}{3\pi^2}\fr{[7!6!]^2}{16!}
\left( \fr{4\pi}{\alpha_s}\right)^{12}
\left( 0.510\,
P\!\! \int_0^{\infty}
dt\fr{t^{N+12}}{1-t} \exp{(-\fr{4\pi}{\alpha_s}t)}+
0.0182 \fr{4\pi}{\alpha_s} e^{-\fr{4\pi}{\alpha_s}}
\right) \,\,\, . \nonumber
\end{eqnarray}
Here, for simplicity, we put  $\phi = 1$ in the principal value
integral (\ref{Pi2}). The value of $N$ should be large enough so
that the perturbative series at $n=N$ will be consistent with the
asymptotics (\ref{asymP}). It is this issue that is most crucial
for the practical applications of our method. As one can see, the
asymptotic prediction (\ref{asymP}) turns out to be many times
larger than the "experimentally known" first $2-3$  terms of the
perturbative series for $\beta$-function or $R_{e^+e^-
\rightarrow hadrons}$. Of course, sooner or later, the series
will reach the asymptotics (\ref{asymP}). However, if it happens
only at $n> 10$ (and it is presumably the case), the issue
of the comparison with the "experiment" will be only of academic
interest.  Moreover at $n> 10$ the renormalons (\ref{uren}),
(\ref{iren}) also may become important.

The only consistent way to find the extent of validity of the
formula (\ref{asymP}) is to consider the $1/n$ corrections to
the leading asymptotics.  The analogous calculation for the
double well oscillator has been performed in our paper
\cite{Faleev}. It is worth mentioning that the accuracy of the
asymptotic formula for oscillator after taking into account of
the  $\sim 1/n$ corrections turns out to be not worse than
$12\%$ starting from $n=5$.

Even for simple quantum-mechanics the calculation of the $1/n$
corrections to the instanton induced asymptotics requires
considerable efforts. The corrections of the same order of
magnitude arose while one takes into account the corrections to
the instanton--anti-instanton interaction (both classical $\sim
e^{-2T}/g^2$ and quantum $\sim e^{-T}$) and the two-loop
correction  $\sim g^2$ to the single instanton density.  The
complete calculation of the similar corrections in QCD seem does
not possible. Nevertheless, the corrections
may be found which have some additional enhancement.  Firstly,
these are the corrections of the order of $\sim ln(n)/n$. In all
formulas for the polarization operator $\Pi (q^2)$ we studiously
kept the function $\phi(x)$ (\ref{eq:phi}) which takes into
account at the two-loop level the variation of the coupling
constant $\alpha_s (\rho_A)$. As we have shown above the use of
the function $\phi(x)$ is necessary for the correct calculation
of the nonperturbative $\sim e^{-\fr{4\pi}{\alpha_s}}$
correction to $\Pi (q^2)$ (\ref{Pi2}). On the other hand,
expanding  $\phi(x)$ in the powers of  $\alpha_s$ one gets the
$\sim [\alpha_s \ln(1-t)]^k/(1-t)$ corrections to the integrand
(\ref{Pi2}).  Each next term of this series contains the small
factor $\alpha_s$ but also the more strong singularity due to
additional $\ln(1-t)$.  It is easy to show that after
integration over $t$ each factor $[ \alpha_s
\ln(\fr{1}{1-t})]^k$ leads to $(\ln(n)/n)^k n!$.  The asymptotic
formula (\ref{asymP}) with all the $\sim( \ln(n)/n)^k$ included
takes the form:
\begin{eqnarray}\label{asymP1}
\Pi_n &=& - 0.156 \left( 1-\fr{3N_c\ln(n)}{n} \right)^{\fr{5}{6}N_c
-\fr{1}{6N_c}} (n+11)! = \\
&=& - 0.156 \left({n}\right)^{-\fr{5N_c^2-1}{2n}} (n+11)!
\,\,\,\, . \nonumber
\end{eqnarray}
As we have said in the introduction, the asymptotic series
(\ref{asymP}) at very large $n$ behave as $n^{11} n^n$. In
(\ref{asymP1}) we have shown explicitly that taking into account
of the asymptotic freedom for the large anti-instanton
($\phi(x)$) leads to additional factor $n^{\fr{1}{n}}$.

We wrote explicitly the factor $5N_c^2$ in the argument
of the exponent in (\ref{asymP1}) in order to show that,
unfortunately, the correction to the leading asymptotics is of
the order of $N_c^2/n \sim 10/n$.  We may also search for the
corrections of the order of $\sim N_c^2/n$  enhanced by the
additional "large" factor $\ln(b)=\ln(3N_c)$. The correction
$\sim N_c^2\ln(n)/n$ (\ref{asymP1}) arose after the function
$\alpha_s(\rho_A)$ was treated consistently at large $\rho_A$.
However, (see discussion after formula (\ref{Kint})), the size of
the "small" instanton also is parametrically large compared to
the characteristic wave length  $\rho_I \sim b/q$. Taking into
account this effect leads to the additional factor in
(\ref{pol.op2}):
\bq\label{dop.fak}
\bigg(\fr{\ln{(\fr{1}{\rho_I
\Lambda_{QCD}})}}{\ln{(\fr{q}{\Lambda_{QCD}})}}
\bigg)^{4N_c-b'/b}=\bigg(1-
\fr{\alpha_s(p)}{4\pi}2b\ln{(b)}\bigg)
^{4N_c-b'/b}.
\ee
Finally, we have:
\bq\label{asymP2}
\Pi_n = - 0.156
\left({(3N_c)^4 n}\right)^{-\fr{5N_c^2-1}{2n}}
(n+11)!
\,\,\,\, .
\ee
Of course, this result is valid only at $n\gg N_c^2$.
Nevertheless it may happen that $n\sim N_c^2 \ln(3N_c)$.

Some other $\sim 1/n$ corrections also are easy to find.  For
example, one may calculate the correction induced by the quantum
part of the instanton--anti-instanton interaction. In fact, in
the leading approximation this correction will be taken into
account if one replaces in the dipole-dipole interaction
$S_{int} = -\fr{4\pi}{\alpha_s} \xi h$ (\ref{action}) the
running constant $\alpha_s(q)$ by the
$\alpha_s(\sqrt{\rho_I \rho_A})$.

However, the complete calculation of all $\sim 1/n$ corrections
is too difficult, and not so actual. As we see the corrections to
the leading asymptotics behave like $N_c^2/n$ and, thus, the
issue of quantitative comparison of the asymptotic prediction
with the exactly calculated terms of the perturbative expansion
will be possible only at $n > N_c^2$.

\section{$R_{e^+e^-\rightarrow hadrons}$}   \label{sec:R}

Now we are able to calculate the instanton contribution to the
cross section of the electron-positron pair annihilation into
hadrons. We have already shown the well known formula (\ref{Pi})
connecting $R_{e^+e^-\rightarrow hadrons}$ and imaginary part of
the polarization operator. The main $q^2$-dependence of $\Pi$
comes from the factor $\exp(-\fr{4\pi}{\alpha_s}) \sim q^{-2b}$.
However, at integer $b$ (in particular if $N_f=N_c$) the
$q^{-2b}$ term does not give rise to the imaginary part. The
imaginary part does appear after the more gentle dependence is
taken into account $\Pi(q^2) \sim
[\ln(\fr{q^2}{\Lambda^2})]^{\gamma} q^{-2b}$. Let us note that
because the imaginary part of the large logarithm $Im\ln(-q^2) =
\pi$ is of the order of one, the $R_{e^+e^-\rightarrow hadrons}$
at $N_f=N_c$ acquires an additional power of $\alpha_s$
compared to the polarization operator $\Pi_{\mu\nu}$.  After
analytical continuation of the expression (\ref{Pi2}) one gets:
\begin{eqnarray}\label{R1}
R_{e^+e^-\rightarrow hadrons}(q^2)&=&
- 11{e^{8/3}}\fr{[7!6!]^2}{15!}
\left( \fr{4\pi}{\alpha_s}\right)^{11}
\left( 0.510
P\! \int_0^{\infty}
dt\fr{\exp{(-\fr{4\pi}{\alpha_s}t)}}{1-t} \psi(|1\!-t|)
\right.  \nonumber \\
&+& \left.
\fr{636}{28985} \fr{4\pi}{\alpha_s} \exp({-\fr{4\pi}{\alpha_s}})
\right) \,\,\,\,\, ,
\end{eqnarray}
where
\bq\label{psi}
\psi(x)=\left( 1+\fr{9}{2}\fr{\alpha_s}{4\pi}\ln(x)\right)
\left( 1+9\fr{\alpha_s}{4\pi}\ln(x)\right)^{\fr{13}{9}}
\,\,\,\,\, .
\ee

Also, it is easy to find the asymptotics of the perturbation
theory together with the $\sim 1/n$ corrections discussed in the
end of preceding section:
\begin{eqnarray}\label{asymR}
R_{e^+e^-\rightarrow hadrons}(q^2)=\sum R_{n}
\left(\fr{\alpha_s}{4\pi}\right)^n
\,\,\, &,& \,\,\, \\
R_{n} =-11 e^{8/3} \fr{[7!6!]^2}{15!} 0.510
\left({9^4 n}\right)^{-\fr{35}{2n}}(n+10)! &\approx&
-813 \left({6561n}\right)^{-\fr{35}{2n}}(n+10)! \,\,\, .
\nonumber
\end{eqnarray}
As we have already said, the "experimentally" known first terms
of the perturbation theory series $R_n$ are in many orders of
magnitude smaller than the naive asymptotics $(n+4N_c)!$. As
well as for the $\Pi_n$ (\ref{asymP2}) the $\sim 1/n$
corrections to $R_n$ (\ref{asymR}) tend to decrease the
asymptotic prediction.  Furthermore, if one extends the
expression (\ref{asymR}) down to $n \sim 1$, this suppression
may even compensate the huge factor $10!$. Let us remind,
however, that one may believe the formula (\ref{asymR}) only at
$n>10$ (or even at $n\gg 10$).  Moreover, deriving the formulas
(\ref{asymP2},\ref{asymR}) we have assumed the following
hierarchy of the $\sim 1/n$ corrections:   $N_c^2\ln(n)/n \gg
N_c^2\ln(N_c)/n \gg N_c^2/n$. But actually, as it is seen from
(\ref{asymR}), even at $N_c=3$ the corrections  $\sim
N_c^2\ln(N_c)/n$ are much more important than the corrections
$\sim N_c^2\ln(n)/n $ for any reasonable $n$. In such situation
one also may expect a surprise from unknown  $\sim N_c^2/n$
corrections.

Up to now we have not considered the question at what energies
our asymptotic formulas may be used. Firstly, we calculate the
functional integral by the steepest descent method. Hence, at
least the coupling constant $\alpha_s(\rho)$ should be small or,
in other words, $\rho \ll 1/\Lambda$.  Moreover we consider only
the case $N_f=N_c$, and so, the minimal value of $\rho$ is
determined by the mass of $c$ quark $\rho_{min} \sim
\fr{1}{m_c}$. On the other hand, as we have shown above, the
effective inverse size of the "small" instanton turns out to be
sufficiently smaller than the external momentum $q \sim
b/\rho_I$. Hence, it is natural to expect that our asymptotic
formulas are valid  within the energy region:
\bq\label{usl}
3Gev \ll qc < 15Gev\ .
\ee

Everywhere above by the coupling constant $\alpha_s(q)$ was
meant the three-flavors coupling  $\alpha_{3}(q)$. However, at
$q$ larger than $5Gev$ it will be natural to express the result
in terms of $\alpha_{5}(q)$. The relation between
$\alpha_{3}(q)$ and $\alpha_{5}(q)$ in the leading approximation
reads:
\bq
 \fr{4\pi}{\alpha_{3}(q)}=\fr{4\pi}{\alpha_{5}(q)}+\fr{4}{3}
\ln{\bigg(\fr{q^2}{m_cm_b}\bigg)}+O(1) \ ,
\ee
where $m_c$ and  $m_b$ are masses of $c$ and $b$ quarks.
As a result, instead of formula (\ref{R1}) we get
\bq\label{R3}
R_{e^+e^-\rightarrow hadrons}(q^2)
=
Const\, \bigg(\fr{m_c m_b}{q^2}\bigg)^{4/3}
\bigg(\fr{4\pi}{\alpha_{5}}\bigg)^{11} P\!\int_0^{\infty} dt
\fr{\exp{(-\fr{4\pi}{\alpha_5}t})}{1-t}
\ee
Here we do not account for any  $\sim 1/n$ corrections
and also do not write very small at $qc \sim 5 Gev$
nonperturbative correction $\sim 1/q^{18}$.

\section{The calculation of the $\tau$ decay width}\label{sec:2}

During the last few years a question about the possibility to
extract the value of the strong coupling $\alpha_s(m_{\tau})$
from the hadronic width of the $\tau$-lepton has been actively
discussed \cite{Pich1}-\cite{Aleph}.  It has been shown that the
"standard", obtained in the framework of QCD sum rules
nonperturbative corrections to the $\tau$ decay width, do not
exceed a few percents \cite{Pich1}. The assumption that only
first terms $\sim 1/q^n$ are important allows one to fix the
value of the strong coupling at the $m_{\tau}$ pole within  10\%
accuracy.

The first attempt to go beyond the "standard" sum rules was made
in the paper of M.Porrati and P.Nason \cite{Nason}.  However,
the obtained result seems to be of only a methodical interest.
The authors of \cite{Nason} have found the single-instanton
contribution in an empty (perturbative) vacuum and consequently
obtained a completely negligible result proportional to the
product of the light quark masses $m_u m_d m_s$.

Much more reasonable approach to the calculation of the
instanton contribution to $R_{\tau \rightarrow hadrons}$ has
been demonstrated in the work of I.Balitsky et al.
\cite{Balitsky1}. They have tryed to take into account the
influence on the instanton of the long-wave nonperturbative
vacuum fluctuations.  Roughly speaking, in the paper
\cite{Balitsky1} the current quark masses were substituted by
the effective masses \cite{Vainshtein3} $m_q \rightarrow m_q
-\fr{2}{3}\pi^2\big<\bar{q}q\big> \rho_I^2$.  Therewith the
instanton contribution turns out to be comparable with the
"standard" nonperturbative contributions.  However, the authors
of \cite{Balitsky1} still do not put in doubt the possibility to
extract the value of $\alpha_s$ from  $R_{\tau \rightarrow
hadrons}$.

In this section we would like to consider the contribution of
the instanton--anti-instanton  pair to the $\tau$ decay width.
By considering the concrete topologically trivial configuration,
we, unlike the authors of the papers \cite{Nason,Balitsky1}, are
able to find the asymptotics of the perturbation theory. As for
the nonperturbative contribution, at first sight one may expect
that our "large" anti-instanton is only one of the examples of
the long-wave vacuum fluctuations which give rise to the quark
condensate $\big<\bar{q}q\big>$ and, hence, our nonperturbative
correction has been already included in the result of the work
\cite{Balitsky1}.  It would be actually the case if our universe
was built only from one or two light quarks ($N_f <N_c$). In
this case the integral over the size of the large anti-instanton
for the polarization operator (\ref{pol.op},\ref{pol.op3})
diverges and it would be natural to evaluate it via some
phenomenological quark condensate $\big<\bar{q}q\big>
\sim \Lambda_{QCD}^3$. However, as we have seen in two preceding
sections, in the case $N_f = N_c$ the integral over $\rho_A$
with the logarithmic accuracy comes from the whole region $1/q
\ll \rho_A \ll 1/\Lambda_{QCD}$.  In particular, it means that
we have found explicitly the most probable long-wave background
for the small instanton. As we will see bellow, the
instanton--anti-instanton pair induced correction to $R_{\tau
\rightarrow hadrons}$ turns out to be much larger than the
semi-phenomenological result of the paper \cite{Balitsky1}.

The reliability of our result also requires the separate
consideration.  The typical size of "small" instanton turns out
to be $b \sim 10$ times larger than $1/q$. Applied to the $\tau$
decay it means that the typical $\rho$ is of the order of
$b/m_{\tau} \sim 1/\Lambda_{QCD}$ to say nothing about the
"larger" anti-instanton. In such situation we may obtain only
more or less reliable estimate of the effect on the order of
magnitude.

Like $R_{e^+ e^- \rightarrow hadrons}$, the ratio of hadronic
$\tau$ decay width to its leptonic width  $R_{\tau \rightarrow
hadrons}$ may be found by the analytical continuation of the
correlator of the weak currents from the euclidean $q^2$ region
(see e.g. \cite{Nason,Balitsky1}):
\begin{eqnarray}\label{Rt}
&\,&R_{\tau \rightarrow hadrons}=-6i\pi\oint_{|s|=1}ds
(1-s)^2[(1+2s)\Pi^T(-sm_{\tau}^2)+
\Pi^L(-sm_{\tau}^2)]\ , \\
&\,&\Pi_{\mu\nu}(q^2)=\Pi^T(q^2)(q_{\mu}q_{\nu}-q^2\delta_{\mu\nu})
+\Pi^L(q^2)q_{\mu}q_{\nu}=
\int dx e^{iqx} \big<j^+_{\mu}(x)j_{\nu}(0)\big>\ , \nonumber
\end{eqnarray}
where
\bq\label{tok}
j_{\mu}=V_{ud}u^+\gamma_{\mu}(1+\gamma^5)d+V_{us}u^+\gamma_{\mu}
(1+\gamma^5)s\ .
\ee
Here the first integral is taken from upper to lower edge of the
cut which goes along the positive real axis of the complex
$s$-plane.  In order to calculate the correlator (\ref{Rt}), one
may use the formulas of the preceding sections with minimal
modifications. The first distinction is that the weak current
(\ref{tok}) include the product of two quark operators with
different flavors. Hence, the nonconnected graphs which in the
case of the correlator of two electro-magnetic currents
(\ref{pol.op}) led to the term proportional to $\big(\sum
e_f\big)^2$,  do not appear  in (\ref{Rt}) (note, in the case of
electro-magnetic currents the term $\sim \big(\sum e_f\big)^2$
vanishes at  $N_f = N_c=3$ due to vanishing of the sum of quark
charges). The fermionic Green function (\ref{GF}),(\ref{GI})
anticommute with $\gamma^5$. As a result, the vector-vector and
axial-axial contributions to the correlator (\ref{Rt}) are
equal. At last, one has to make the substitution $\sum_f
e^2_f(=2/3) \rightarrow (|V_{ud}|^2+|V_{us}|^2)(\simeq 1)$ while
going from (\ref{pol.op}) to (\ref{Rt}).  Note that both
correlators (\ref{pol.op}) and (\ref{Rt}) turn out to be
transverse. Thus, we have $\Pi^T(q^2)=3\Pi(q^2)$ (see.
(\ref{Pi2})) and $\Pi^L(q^2)=0$. Now, integrating over $s$, one
can easily find $R_{\tau \rightarrow hadrons}$ in the leading
over $\alpha_s$ approximation
\begin{eqnarray}\label{Rt1}
&\,& R_{\tau \rightarrow
hadrons}=\fr{33e^{8/3}}{40}\fr{[7!6!]^2}{15!}
\left( \fr{4\pi}{\alpha_s}\right)^{11}
\left( 0.510
 P\int_0^{\infty}
dt\fr{\exp{(-\fr{4\pi}{\alpha_s}t)}}{1-t} \psi(|1\!-t|)
\right.
\\  &\,& \quad \quad \quad \quad \left. +
\fr{636}{28985} \fr{4\pi}{\alpha_s}
e^{-\fr{4\pi}{\alpha_s}}\right) \, . \nonumber
\end{eqnarray}
This expression leads to the following asymptotics of the
perturbation theory :
\begin{eqnarray}\label{Rntau}
R_{\tau \rightarrow hadrons}&=&\sum R_{n\tau}\left(
\fr{\alpha_s}{4\pi}\right)^{n} \\
R_{n\tau}&=&
61(n+10)!\bigg(9^4n\bigg)^{-\fr{35}{2n}} \ . \nonumber
\end{eqnarray}
Just as for $R_{e^+e^- \rightarrow hadrons}$ (\ref{asymR}),
there appear certain problems while interpreting the
instanton--anti-instanton pair induced asymptotics of the
perturbation theory for $R_{\tau \rightarrow hadrons}$ . In
fact, the asymptotics (\ref{Rntau}) is valid only at $n \gg 10$,
where the renormalon contributions (\ref{uren}),(\ref{iren}) are
much larger than the instanton one. We may only hope that at $n
\sim 10$ formula (\ref{Rntau}) does give the correct estimate of
the instanton contribution on the order of magnitude.  If so, at
$n \sim 10$ the instantons will dominate over renormalons. In
the most interesting case $n = 3-5$ our formulas do not work.

In order to compare our result with the experiment, let us
consider in more detail the pure nonperturbative  $\sim
e^{-\fr{4\pi}{\alpha}}$ term in (\ref{Rt1}) (note that this
correction in terms of $\Lambda_{QCD}$ (\ref{eq:alpha}) behaves
like $(\Lambda/m_{\tau})^{18}$). The quantities $R_{\tau}^{np}$
for popular values of $\alpha_s(m_{\tau})$ are shown in the
first column of the Table.  One has to compare these values with
the experimental value $R_{\tau \rightarrow hadrons}=3.56
\pm 0.03$ \cite{Pich3} (moreover, the last quantity contains a
large trivial part $R_{\tau \rightarrow hadrons} \approx N_c$
and one has to extract the value $\alpha_s$ only from the
remainder $R_{\tau \rightarrow hadrons}-3 = 0.56 \pm 0.03$).  As
we can see, the nonperturbative correction turns out to be
dramatically large.

There exist, however, the procedure (used also in the work
\cite{Balitsky1}) which allows to reduce the huge discrepancy
with the experiment. The regular way to improve the
nonperturbative correction (\ref{Rt1}) is to calculate the $\sim
\alpha_s$ corrections to it. Since, as will be shown bellow,
the result is going to be deceased by $30-50$ times, one has to
sum an infinite series of the corrections. Undoubtedly, the exact
calculation even of the first correction of the order of
$\alpha_s$ to (\ref{Rt1}) is beyond our abilities.  All we can
do is to use the dependence of the coupling constant on the
instanton size $\alpha_s(\rho)$, which is known from the
renorm-invariance principle.  Rather weak justification for
taking into account just these particular corrections is the
fact that they are enhanced as $\ln(3 N_c)$ compared to the
other ones.  As we have said above, the typical inverse size of
the "small" instanton is much smaller than the external momentum
$(\rho^*)^{-1} \sim m_{\tau}/b$. The dependence of the coupling
constant on the instanton size in the leading one-loop
approximation has been already taken into account when we wrote
the instanton density in the form $d(\rho)\sim \rho^b$. At first
sight, in order to account for the two-loop effects, one has
simply to add to the nonperturbative contribution (\ref{Rt1})
the factor (see (\ref{dop.fak})) :
\bq\label{loglog}
\left( \fr{4\pi}{\alpha_s} \right)^{13} e^{-\fr{4\pi}{\alpha_s}}
\rightarrow \left(
\fr{\ln\left(\fr{m_{\tau}}{b\Lambda}\right)
}{\ln\left(\fr{m_{\tau}}{\Lambda}\right)}\right)^{\fr{53}{9}}
\left( \fr{4\pi}{\alpha_s} \right)^{13} e^{-\fr{4\pi}{\alpha_s}}
\ .
\ee
Unfortunately, just for $\tau$-lepton the formula (\ref{loglog})
turns out to be very unstable.  Deriving  (\ref{loglog}), we
have implied that $\ln\left(\fr{m_{\tau}}{b\Lambda}\right) \gg
1$, while actually $m_{\tau} \approx b\Lambda$. In order to
improve the situation, we will try to insert the exact function
$[-\ln(\rho\Lambda)]^{\fr{53}{9}}$ in the $\rho$-integral
(\ref{Kint}), as it was done in the paper
\cite{Balitsky1}.  While doing so, we effectively shift the maximum of
the integrand from the dangerous region $\rho \sim 1/\Lambda$
for $53/9 \approx 6$ is sufficiently large quantity.  As the
result, the main contribution to  $R_{\tau \rightarrow
hadrons}^{np}$ comes from the instantons of the size $\rho \sim
4/m_{\tau}$ rather than $\rho \sim b/m_{\tau}$ . Integrating
over the  Borel parameter (see (\ref{Pi1})-(\ref{Pi2})) before
the $\rho$-integration (just after the Borel integration the
full multiplier $[\ln(\rho\Lambda)]^{\fr{53}{9}}$ is gathered),
we obtain the following expression for the nonperturbative
correction to  $\Pi (q^2)$:
\begin{eqnarray}\label{p-s}
&\,&\Pi^{np} (-s m_{\tau}^2) = \fr{e^{8/3}}{\pi^2 \cdot 7 \cdot
9 \cdot
31 \cdot 2^{18}} \left( \fr{4\pi}{\alpha_s(m_{\tau})}\right)^{13}
\exp{\big(-\fr{4\pi}{\alpha_s(m_{\tau})}\big)} \times \\
&\,& \quad \int_0^1 du \int \fr{dx}{(-s)^9} x^{15} \left\{
\fr{1}{2u\overline{u}}
K_2\left(\fr{x}{\sqrt{u\overline{u}}}\right)
-\fr{7}{5}K_0\left(\fr{x}{\sqrt{u\overline{u}}}\right) \right\}
\left( 1+ \fr{9\alpha_s}{4\pi}
\ln\left(\fr{-s}{x^2}\right)\right)^{\fr{53}{9}} \ . \nonumber
\end{eqnarray}
Here we have also performed the analytical continuation
(\ref{Rt}) $m_{\tau}^2 \rightarrow -s m_{\tau}^2$ . In order to
calculate the integral over $x$, one may use the identity:
\bq\label{vareps}
\ln(p)^n = \lim_{\varepsilon \rightarrow 0} \left(
\fr{\partial}{\partial\varepsilon} \right)^n p^{\varepsilon} \ .
\ee
Now, with the use of (\ref{Rt}), we can find the nonperturbative
correction to the $\tau$ decay width
\begin{eqnarray}\label{Rtau1}
 R_{\tau \rightarrow hadrons}^{*np}&=& \lim_{\varepsilon
\rightarrow 0} \, -i\fr{e^{8/3}}{\pi \cdot 7 \cdot 31
\cdot 2^{17} }
\left( \fr{4\pi}{\alpha_s(m_{\tau})}\right)^{13}
\exp{\big(-\fr{4\pi}{\alpha_s(m_{\tau})}\big)}
\times \\ &\,& \left( 1+ \fr{9\alpha_s}{4\pi}
\fr{\partial}{\partial \varepsilon} \right)^{\fr{53}{9}}
 \int ds \fr{1-3s^2+2s^3}{(-s)^{9-\varepsilon}} dx\,
x^{15-2\varepsilon} du \left\{
\fr{K_2}{2u\overline{u}}
-\fr{7}{5}K_0 \right\} \ . \nonumber
\end{eqnarray}
After trivial integration over $x, s$ and $u$ one gets:
\begin{eqnarray}\label{Rtau2}
R_{\tau \rightarrow hadrons}^{*np}&=& \lim_{\varepsilon
\rightarrow 0} \, \fr{9 e^{8/3}}{\pi \cdot 70 \cdot 31}
\left( \fr{4\pi}{\alpha_s(m_{\tau})}\right)^{13}
\exp{\big(-\fr{4\pi}{\alpha_s(m_{\tau})}\big)} \\ &\times&
\left( 1+ \fr{9\alpha_s}{4\pi}
\fr{\partial}{\partial \varepsilon} \right)^{\fr{53}{9}}
\sin(\varepsilon\pi) \fr{(12-\varepsilon)\Gamma(9-\varepsilon)
\Gamma(8-\varepsilon)^2
\Gamma(5-\varepsilon)}{2^{2\varepsilon}\Gamma(18-2\varepsilon)}
\ . \nonumber
\end{eqnarray}
This expression also may be presented as a series in powers of
$\alpha_s$ . Note, that the first term of the series,
corresponding to $({\partial}/{\partial\varepsilon})^0$, is
vanishing.  This leads to loss of one factor
$1/\alpha_s(m_{\tau})$  in $R_{\tau \rightarrow hadrons}$
(similarly, as in the case of $R_{e^+e^- \rightarrow hadrons}$)
compared to the correlation function. Coefficients of the
expansion of (\ref{Rtau2}) in powers of $\alpha_s$ were found
numerically. The resulting quantities  $R_{\tau \rightarrow
hadrons}^{*np}$ are shown in the last column of the Table. The
terms of series in $\alpha_s$, generated by $\left( 1+
\fr{9\alpha_s}{4\pi}
\fr{\partial}{\partial \varepsilon} \right)^{\fr{53}{9}}$, are
maximal at $n\sim2\div3$ and become negligible at $n>8$ . As we
can see, taking into account of the two-loop dependence of
$\alpha_s(\rho)$ allows to reduce the nonperturbative correction
almost by two orders of magnitude. Nevertheless, the formula
(\ref{Rtau2}) turns out to be surprisingly stable. We have used
also the Stirling formula for approximate calculation of the
$\Gamma$-functions in (\ref{Rtau2}) and the approximate values
of $R_{\tau\rightarrow hadrons}^{*np}$ found in this way almost
coincide with results of numerical calculations listed in the
table.

It is seen from the table that passing from $\alpha_s = 0.28$ to
$\alpha_s = 0.29$ our correction changes the sign.  Moreover,
the ratio  $R^*/R$ in the interval  $0.15 < \alpha_s < 0.43$
changes the sign three times, being everywhere smaller than
$1/30$.  Compared to the simple formula (\ref{loglog}) such rich
behaviour provides us one more evidence that effectively we work
at very low energies and listed in the table result is, at best,
only the estimate on the order of magnitude.

\begin{flushright}
{\bf Table}
\end{flushright}
\begin{center}
\begin{tabular}{|c|c|c|}
\hline
$\alpha_s(m_{\tau})$ &$R_{\tau\rightarrow
hadrons}^{np}$ & $R_{\tau\rightarrow hadrons}^{*np}$  \\ \hline
   0.28  &  5.65    & -0.0229   \\ \hline
   0.29  &  17.44   &  0.0507  \\ \hline
   0.30  &  49.23   &  0.476  \\ \hline
   0.32  &  311.1   &  6.70  \\ \hline
   0.34  &  1514.1  &  44.79  \\ \hline
   0.36  &  5943.4  &  190.1  \\ \hline
\end{tabular}

{\small The full nonperturbative contribution of the
instanton--anti-instanton  pair to the hadronic $\tau$ decay
width $R_{\tau \rightarrow hadrons}^{*np}$ and the contribution
accounting for only the leading over $\alpha_s$ term in the sum
(\ref{Rtau2}) $R_{\tau \rightarrow hadrons}^{np}$ at various
values of $\alpha_s(m_{\tau})$. }
\end{center}

One should remember that the obtained results should be compared
with the experimental value $(R_{\tau \rightarrow hadrons}
-3)_{exp} = 0.56 \pm 0.03$ .  We see that, in spite of all our
effort, even for  $\alpha_s = 0.29$ the instanton contribution
to $R_{\tau \rightarrow hadrons}$ still remains large. In this
situation the only way out may be to ignore completely the
instanton contribution to  $R_{\tau \rightarrow hadrons}$ . For
example, one may say that the series of the power corrections
$\sim 1/m_{\tau}^n$ is also asymptotic and it is natural to cut
off it somewhere at $n \sim 4-8$ (the instanton contribution
behaves like $\sim 1/m_{\tau}^{18}$). Nevertheless, we do not
know to what extent this point of view is justified and we
consider our result as indication of the  {\it impossibility} to
extract the reliable value of $\alpha_s$ from $R_{\tau
\rightarrow hadrons}$ .

Authors are thankful to V.~L.~Chernyak, M.~E.~Pospelov,
A.~I.~Vainshtein and A.~S.~Yelkhovsky for valuable discussions.
This work was supported by the Russian Foundation for
Fundamental Research under Grant 95-02-04607a. The work of S.F.
has been supported by the INTAS Grant 93-2492 within the program
of ICFPM of support for young scientists.

\section{Appendix. Integration over the $SU(3)$ group }

The average values of  expressions polynomial in the elements of
the $SU(N)$ matrix may be easily found by, for example,
graphical method described in \cite{Creutz}. However, for
calculation of the integrals including  $\theta$-functions like
$\int |TrO|^6\theta(h)$ (\ref{pol.op3}) one has to specify in
explicit form the parametrization of the matrix and the measure
of integration. We use the following parametrization of the
$SU(3)$ group element:
\begin{eqnarray}\label{matr}
U\!=\!\left(\begin{array}{ccc}
-P_2^* P_4^* s_1s_2-P_1P_3P_5^* c_1c_2c_3 &
P_1^* P_4^* c_1s_2-P_2P_3P_5^* s_1c_2c_3 & P_3c_2s_3 \\
P_2^* P_3^* s_1c_2-P_1P_4P_5^* c_1s_2c_3 &
-P_1^* P_3^* c_1c_2-P_2P_4P_5^* s_1s_2c_3 &  P_4s_2s_3 \\
P_1c_1s_3  & P_2s_1s_3 &                P_5c_3 \\
\end{array} \right) ,
\end{eqnarray}
where: $c_{1,2}=\cos\fr{\psi_{1,2}}{2}\, ,\
s_{1,2}=\sin\fr{\psi_{1,2}}{2}\,, c_3=\cos\psi_3\,$ and
$s_3=\sin\psi_3\,,\ P_n=e^{i\varphi_n}$.  The parameters
$\psi_m$ and $\varphi_n$ vary in the range: $0<\psi_{1,2}<\pi
\,, \ 0<\psi_3<\pi/2$ and $0<\varphi_n<2\pi$.  It may be shown
that the matrix (\ref{matr}) is unitary and unimodular. The
measure of integration over the matrix group $U$ is given by:
\begin{eqnarray}\label{mera}
 dU =N\int \Pi\alpha_i|\det{(M)}|^{1/2} \ , \\
M_{ij}=Tr(U^{-1}(\partial_iU)U^{-1}(\partial_jU))\ . \nonumber
\end{eqnarray}
In our case the parameters $\alpha_i$ ($i=1,...,8$)
are the three angles $\psi_m$ and five angles $\varphi_n$.
After rather tedious algebra one obtaines:
\bq\label{mera1}
 dU=N\Pi d\psi_m \Pi d\varphi_n
\sin\psi_1 \sin\psi_2 \sin^3\psi_3 \cos\psi_3 \ .
\ee
The results of the averaging with this measure of the expressions
$\langle |TrO|^6\rangle $ and \\
$\langle |TrO|^2Re[detO(TrO^+)^2]\rangle$ coincide with those
obtained by the graphical method \cite{Creutz}.
The integration of the expressions including $\theta$-functions
has been performed numerically. The results are given bellow:
\begin{eqnarray}\label{eq:group}
\langle |TrO|^6\rangle  =\fr{7}{5} \, &;& \,
\langle |TrO|^6\theta(h)\rangle =1.368\,
 ;\ \ \ \ \ \ \ \ \
\ \ \ \ \ \nonumber \\
\langle |TrO|^2Re[detO(TrO^+)^2]\rangle =\fr{1}{2} \, &;& \,
\langle |TrO|^2Re[detO(TrO^+)^2]\theta(h)\rangle =0.473\, ;
\nonumber \\
\langle A|TrO|^6\rangle = \fr{48}{85} \,&;& \,
\langle A|TrO|^6\theta(h)\rangle = 0.510 \, .
\end{eqnarray}

\end{document}